\documentclass[preprint]{aastex631}

\makeatletter
\renewcommand{\frontmatter@title@above}{}
\makeatother

\renewcommand*{\thefootnote}{\fnsymbol{footnote}}

\begin{document}

\begin{figure}
\centering
\includegraphics[width=3.5cm]{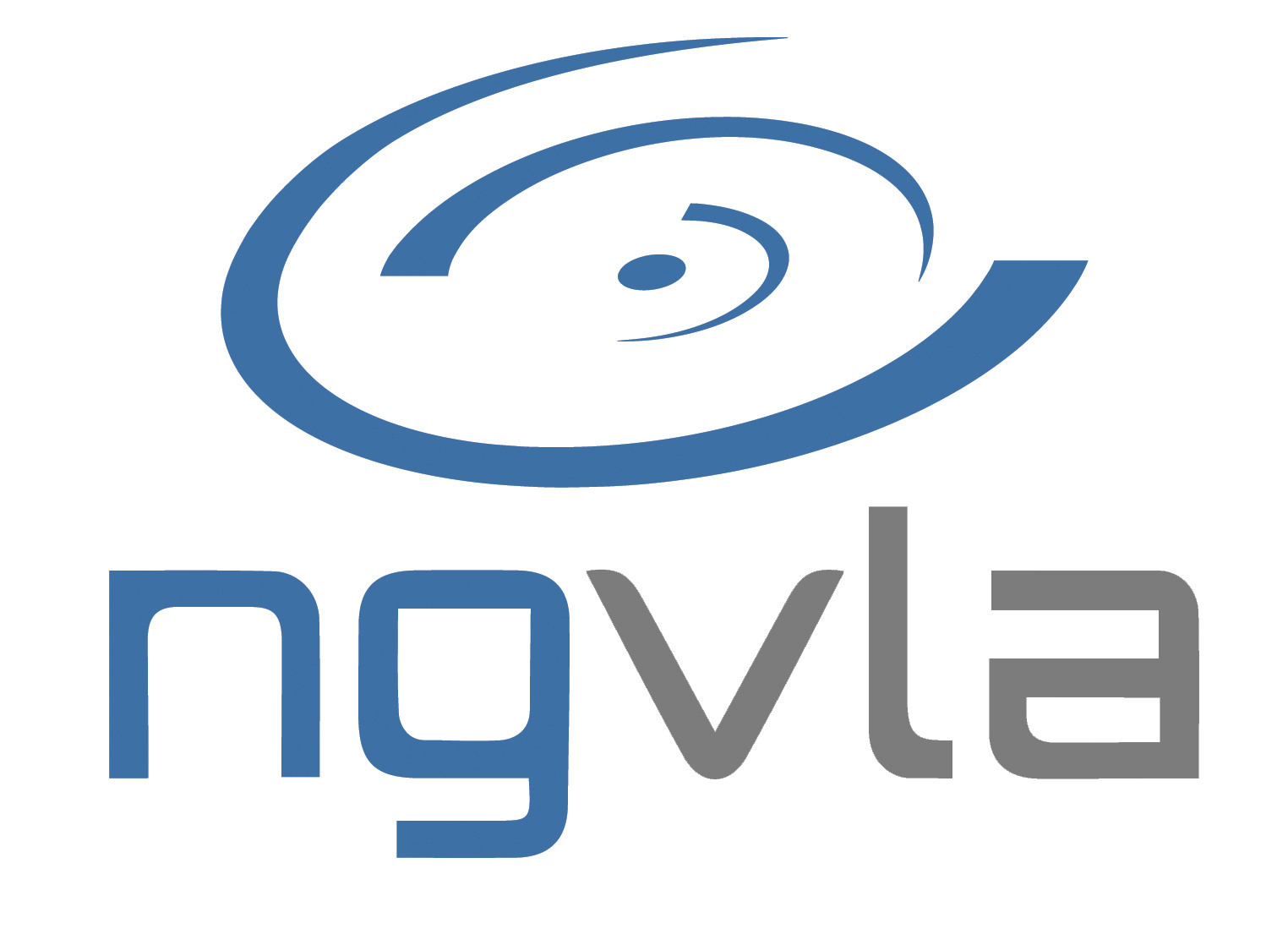}
{\center\Large\bf ngVLA Electronics Memo No. 21 }
\end{figure}

\title{The ngVLA Band-4 Water Vapor Radiometry Concept}
\author{T. K. Sridharan}
\affiliation{National Radio Astronomy Observatory, 520 Edgemont Rd. Charlottesville, VA 22903, USA}

\author{R. Lehmensiek}
\affiliation{National Radio Astronomy Observatory, 520 Edgemont Rd. Charlottesville, VA 22903, USA}
\affiliation{Central Development Laboratory, 1180 Boxwood Estate Road, Charlottesville, VA 22903, USA}

\author{Yoshiharu Asaki}
\affiliation{Joint ALMA Observatory,
    Alonso de C\'{o}rdova 3107, Vitacura 763 0355, 
    Santiago, Chile}
\affiliation{National Astronomical Observatory of Japan,
    2-21-1 Osawa, Mitaka, Tokyo, Japan}    

\pagestyle{plain} 
\pagenumbering{arabic}

%\maketitle
\pagestyle{plain} 
\pagenumbering{arabic}

\begin{abstract}
    The ngVLA has adopted a standalone system with a $\sim$~1.5 m diameter antenna and associated ambient temperature receiver and electronics at each antenna as its baseline design for water vapor radiometer (WVR).  The WVR is intended to decrease calibration overheads for tropospheric phase correction for the high frequency bands (5 \& 6; 30.5-116 GHz) to levels lower than feasible with fast switched reference gain calibration. 
    
    In this memo, we present an alternative concept  utilizing the standard Band-4 science receivers and the main antennas. This would require  the Band-4 feed to be placed between those of Bands 5 \& 6 resulting in a beam offset of 0.8 deg, similar to the beam size of the standalone 1.5 m antenna. The main benefits of the Band-4 WVR concept are: (1) sensing water vapor in a region more representative of the Band 5 \& 6 science beams in the lower layers of the troposphere, in the near-field of the 18 m antenna, than the 1.5 m antenna (2) significantly relaxing the stability requirements due to the lower receiver temperature of the standard Band-4  cryogenic receiver and (3) eliminating the procurement, deployment, and maintenance of a large number of standalone units. Given these advantages, it is recommended that the details and the pros and cons of Band-4 WVR implementation compared a standalone system be carefully considered. \end{abstract}

\section{Introduction \& Motivation} 

The ngVLA science and system requirements call for image dynamic ranges of 32 and 28 dB in Bands 5 \& 6 and set a calibration efficiency goal of 90\% \citep{2021ngVLASysReq}. Achieving adequate phase correction to fulfill these dynamic range targets at the highest frequencies (mm-wavelength Bands~5 and 6) with small calibration overheads is challenging. Tropospheric delay variation on short time scales due to the fluctuating water vapor column above each antenna is the main source of interferometric visibility errors in these bands, and therefore, image dynamic range degradation. Fast-switching to a nearby reference calibrator is the standard approach to correct for such fluctuations at mm and sub-mm wavelengths, which provides an obvious default strategy for ngVLA.  The high sensitivity of ngVLA which affords the ability to find bright enough calibrators needing small integration times within small angular distances \citep{2015ngVLAMEM001, 2022ngVLAMEM98}, makes this approach even more viable. 

However, due to the short timescales of tropospheric fluctuations, delivering the dynamic range targets requires similarly short calibration cycle times, leading to low observing efficiency. For a \mbox{30~s} switching cycle, the predicted nominal dynamic range for 107 antennas in a 1 hr observation is estimated to be $\sim$ 28 dB at 116 GHz, using expected residual phase errors after calibration under the best month nighttime median conditions, which meets the dynamic range requirement (\cite{DraftCalReq} and phase fluctuation estimates in \cite{Asaki2025ngVLAMEM135}, Fig 1; based on data in \cite{1999VLAMEMO222}). Given the slew rate and settling time performance specifications of the ngVLA prototype antenna, the short cycle times required entail a calibration efficiency of $\sim$ 50\% or less, falling well short of  the desired on-target observing efficiency goal of 90\%. If amplitude loss from decoherence is also considered, although this is not an ngVLA requirement, the necessary calibration cycle timescales become even shorter \citep{Asaki2025ngVLAMEM135} which further lowers efficiency or renders fast switching impractical. Under such low coherence conditions, the ngVLA would abandon high frequency observations and switch to more tolerant lower frequency observations \citep{2024ngVLACalConcept}.

The prospective technique adopted in the baseline ngVLA design to improve the calibration efficiency of tropospheric delay corrections on short time scales (seconds) for  high frequency observations is water vapor radiometry (WVR) (\cite{Clarke2015WVRngVLAMEM10, 2020ngVLAMEM061, Towne2020WVRngVLAMEM79, 2021ngVLASysReq, 2024ngVLASysConDesRep, 2024ngVLACalConcept, Asaki2025ngVLAMEM135}; see earlier work by \cite{1999VLAMEMO177} for extensive considerations and modeling in the context of the VLA). The technique detects the 22 GHz water vapor line emission from the troposphere and uses its parameters to estimate excess delay on short time scales for later correction (\cite{1999VLAMEMO177} includes a good overview and further references to previous work). 

For lower frequencies, the high ngVLA sensitivity allows tracking gain fluctuations of all origins with fine time granularity through self-calibration \citep{2023ngVLAMEM108}. For high frequencies, the ngVLA antenna drive performance for fast-switching to a calibrator up to 3 deg away \citep{2022AntTechReq} can only attain the 90\% observing efficiency goal on 200 s time scales (e.g. \{\mbox{180 s} target integration – 7 s slew \& settle - 6 s calibrator integration – 7 s slew \& settle – target\} .. cycle) which is far too long to meet the dynamic range requirements. Such fast-switching capability is required, however, for continuous recalibration of the weights and scaling factors needed to derive delay corrections from the WVR data (section 4.2). 

The overall implementation agnostic requirements for the measurement accuracy and timescales of ngVLA water vapor radiometry have been derived in \cite{Asaki2025ngVLAMEM135}.

The currently baselined ngVLA WVR implementation strategy calls for a dedicated standalone WVR system on each antenna/site, with a reflector of diameter $\sim$ 1.5~m  and associated receivers and electronics \citep{2022WVRTechReq, 2022WVRDes} - 244 units for the 18 m antennas and up to 6 units with tracking mounts for the 6 m antennas. Such a WVR reflector would have a beam size of $\sim$ 0.8$^{\circ}$ and would be mounted laterally offset up to 18 m\footnote{$\sim$ 10 m in the current concept of the standalone implementation} from the main antenna (\cite{2022WVRTechReq}; e.g. attached to the antenna at its periphery). Given the risk implied by the lack of such small dish standalone WVR systems previously deployed at interferometric radio astronomy observatories and the extensive effort entailed by this implementation approach throughout the life cycle of the ngVLA, it is necessary to evaluate alternatives before selecting a final strategy.  

An obvious alternative is the use of the main ngVLA antennas and the Band 4 science receivers for WVR. The Band 4 receivers feeding the primary antennas span a frequency range of 20.5 – 34 GHz, covering the 22 GHz water line. This memo proposes and documents the Band-4 WVR concept. The pros and cons to clarify the tradeoff between the dedicated, standalone system and the main antenna-based system are discussed.

\section{Off-axis Band-4 WVR Concept}

The proposed concept envisages acquisition of tropospheric water vapor data using the main antenna and the standard Band-4 science receiver during, and simultaneously with, science observations in the other bands requiring WVR corrections, primarily 5 \& 6. It provides continuous monitoring, with very short equivalent cycling times (seconds). The spectra will be constructed by a WVR-dedicated digital processor at each antenna, in the same (or similar) way as for a standalone WVR, with a coarse resolution of $\sim$ 100 MHz. Tracking fluctuations in the delay (differential) is preferred, estimated from the WVR spectra by a suitable algorithm, rather than the total absolute bulk delay \citep{1999VLAMEMO177}, which can also be estimated to lower precision. It is clear that this approach is directly available during Band-4 science observations. 

The primary motivating merits driving the concept are the benefits of using the main antennas and the standard science receivers: (1) the possibility of a more representative sampling of the same delay causing water vapor column as the science observation due to better overlap with the science beam (identical for Band 4) (2) complete elimination of a separate WVR system – its design, development and testing; procurement/construction of a large number of units ($\sim 250$), array wide deployment, alignment, and maintenance and (3) leveraging the effort to deliver high sensitivity and stability for ngVLA science receivers, as opposed to the separate effort needed in delivering high performance in a second system. We recognize that separating the two may offer some advantages and the benefits we foresee in (2) and (3) are subject to specific implementation details and associated costs which are left to the relevant engineering teams for careful evaluation.  We restrict ourselves to the viability and advantages of the Band-4 WVR concept for tropospheric delay  correction in the rest of the document.  In view of (1), we note that all previous and existing WVR systems at interferometric radio telescopes, both operational and experimental (e.g. OVRO, PdBI/NOEMA, ATCA, ALMA and the VLA experimental systems), used the main antenna with an offset WVR front end. In the VLA experimental systems, the VLA K band science front end forms the WVR front end, in the same way as proposed here in the Band-4 concept. 

It is important to recognize that the tropospheric fluctuations originate entirely in the near-field of the 18~m ngVLA antenna for both the science and Band-4 WVR beams while partly in the near- and far- fields of a 1.5 m standalone WVR antenna. As the water vapor content falls off with height, the better matching this implies between the science and the Band-4 WVR beams in the lower layers of the troposphere closer to the antenna is an advantage. %This is not the case for a standalone WVR with a lateral position offset from the primary. 

\begin{figure}
\begin{center}  
\includegraphics[width=0.7\linewidth]{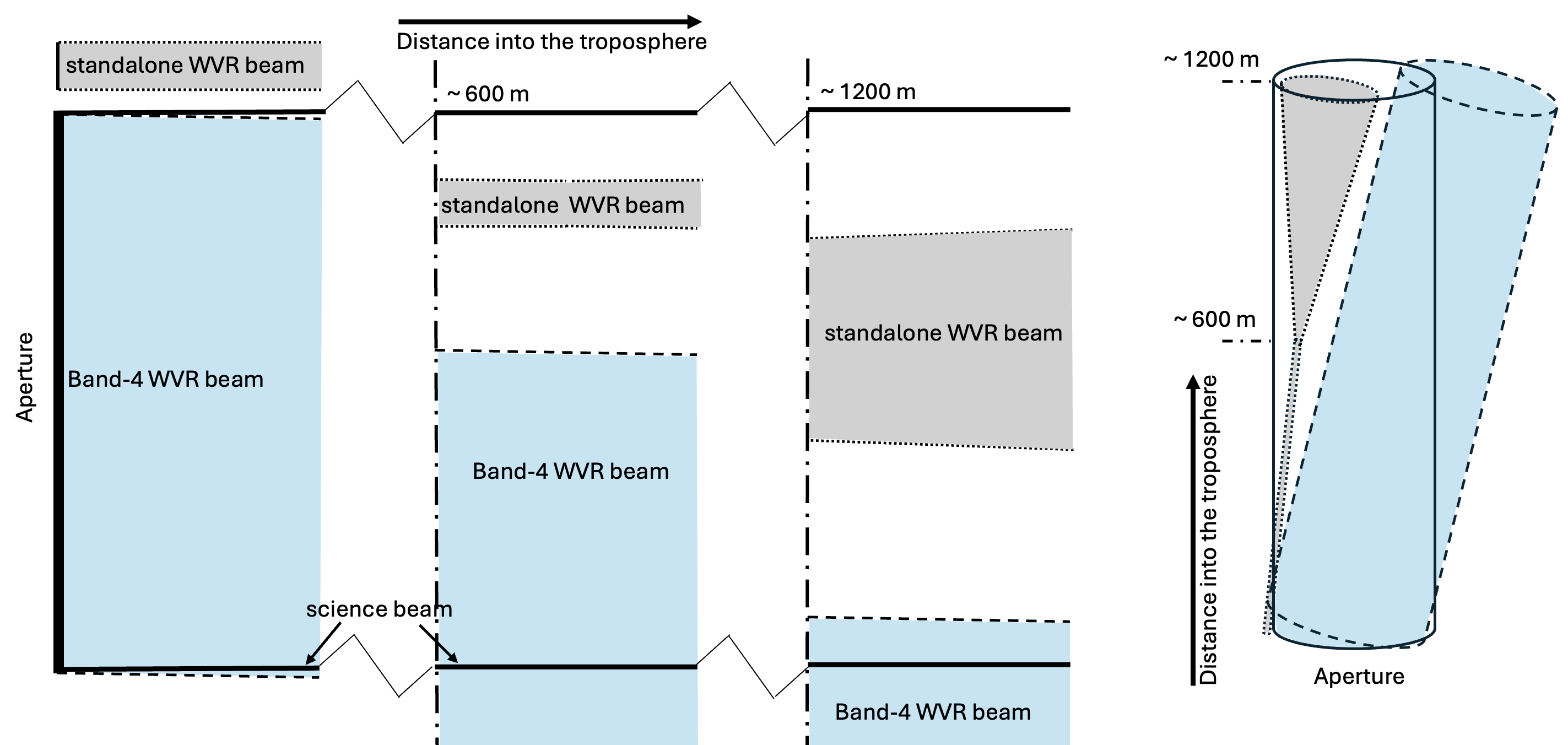}
\caption{\small Schematic representations of the approximate response regions and interrelationships of the standalone WVR (gray areas with dotted boundaries), Band-4 WVR (light blue areas with dashed line boundaries), and the Band 5 \& 6 science feeds (open areas with solid boundaries) in the troposphere. The left panel with three sections depicts the cross sections of the response regions at three heights: the aperture, a representative near-field distance for the 1.5 m standalone antenna of $\sim$ 600 m, and a representative phase screen distance of $\sim$ \mbox{1200~m}, approximately to scale. The right most panel is a schematic 3-D representation, not to scale. The Band-4 WVR response region is more representative the response regions of the science feeds in the lower layers of the troposphere.}
\label{Fig: WVRandScienceBeams}
\end{center}
\end{figure} 

Figure 1 shows a schematic representation of the response regions of the standalone WVR, the Band-4 WVR, and the science feeds in the troposphere. The standalone WVR response can be approximated as a thin cylinder of $\sim$ 1.5 m diameter up to its Rayleigh-Fraunhofer near-field \mbox{distance ($2D^2/\lambda$)} of \mbox{$\sim$~600~m}, laterally offset by up to 18 m\footnote{$\sim$ 10 m in the current concept of the standalone implementation} at its base from the primary antenna science beam, and a cone diverging at the far-field beam angle rate thereafter. For the Band-4 WVR, this distance is 47 km, well outside the troposphere, up to which the response region is a cylinder of diameter 18 m (same as the main antenna) tilted by the offset angle, but coincident with the science beam at its base at the aperture. In addition to water vapor content decreasing with height leading to a larger contribution to the delay from the layers closer to the aperture, note that for small sampling time scales, the lateral offset of the standalone WVR is comparable to the size scales from which the fluctuations originate ($\sim$~10~m for 2 s). This makes the Band-4 WVR beam a better matched probe of the fluctuating water vapor column impacting the science observation. This shortcoming of a standalone WVR has also been pointed out in the context of VGOS \citep{2021JGeod..95..117F}, although without the near-field considerations. A full near-field treatment is in progress and we intend to update this memo with the results in future if warranted (Lehmensiek \& Sridharan, in preparation).

An immediately apparent drawback of the concept, which we look at more carefully in this memo, is that during observations in Bands 5 \& 6 which need WVR corrections, the Band-4 receiver would be offset from the focus, and consequently, its water vapor probing beam would be correspondingly off axis. The viability of the concept depends on how far off-axis the offset Band-4 beam would be and the characteristics of such an off-axis beam, which in turn depend on the layout of the receivers for the different bands (Figure 2; further discussed below).  
\begin{figure}[h!]
%\begin{figure}
\begin{center}   
\includegraphics[width=0.6\linewidth]{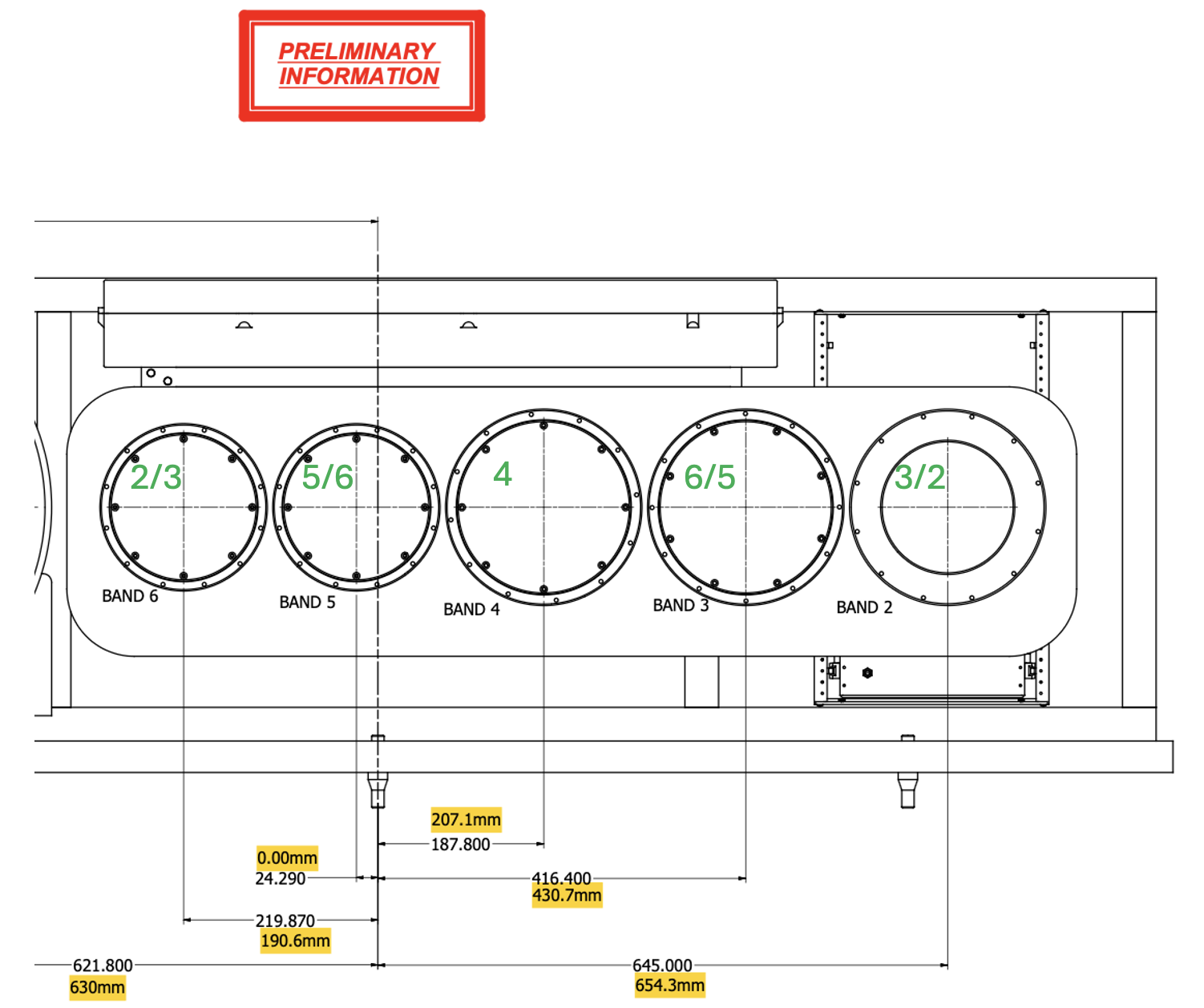}
\caption{\small Preliminary ngVLA frontend layout (dimensions in mm) \citep{Sturgis2021FE_Layout_DWG_Draft}. The distances in the current version of the layout are marked in yellow. The proposed locations for the feeds are labeled in green. Band-4 feed is already at the center.}
\label{Fig: FrontendLayout}
\end{center}
\end{figure}

While this approach does not deliver full overlap between the science beam and the WVR probe beam except during Band-4 science observations, the matching is better than with a $\sim$ 1.5 m standalone system. It also provides a better match than fast switching to a calibrator up to 1-2 deg away and eliminates calibration overheads without having to implement separate WVR system hardware. An exact matching of the science and the probe beams is fundamentally impossible without splitting the science beam in some fashion (spatially in power/polarization or in frequency).

Figure 2 shows the layout of the receiver front ends in the preliminary ngVLA reference front end design \citep{Sturgis2021FE_Layout_DWG_Draft}, excluding Band-1, which is not relevant for this discussion (as it is not expected to need tropospheric water vapor delay correction). Bands 2 \& 3 also do not require WVR corrections \citep{2023ngVLAMEM108}. The front end enclosure is mounted on a linear translation stage (feed indexer) and  an observing band is chosen by moving the feed for that band to the focus position. Clearly, it will be beneficial to place the Band-4 feed at the center of the linear layout with the feeds of the bands with the most stringent need for WVR correction located closest to it on either side, in the sequence (2/3)-(5/6)-4-(6/5)-(3/2) to minimize the 4-6 and 4-5 feed offsets. Based on the layout, the range of distances to these feeds from the Band 4 feed would be $\sim$ 200 – 450 mm. The 450 mm distance to Band 2 \& 3 are considered in order to possibly provide less precise information on bulk tropospheric delay due to water vapor to the observations in these bands. 

\begin{figure}[h!]
%\begin{figure}
\begin{center}   
\includegraphics[width=0.7\linewidth]{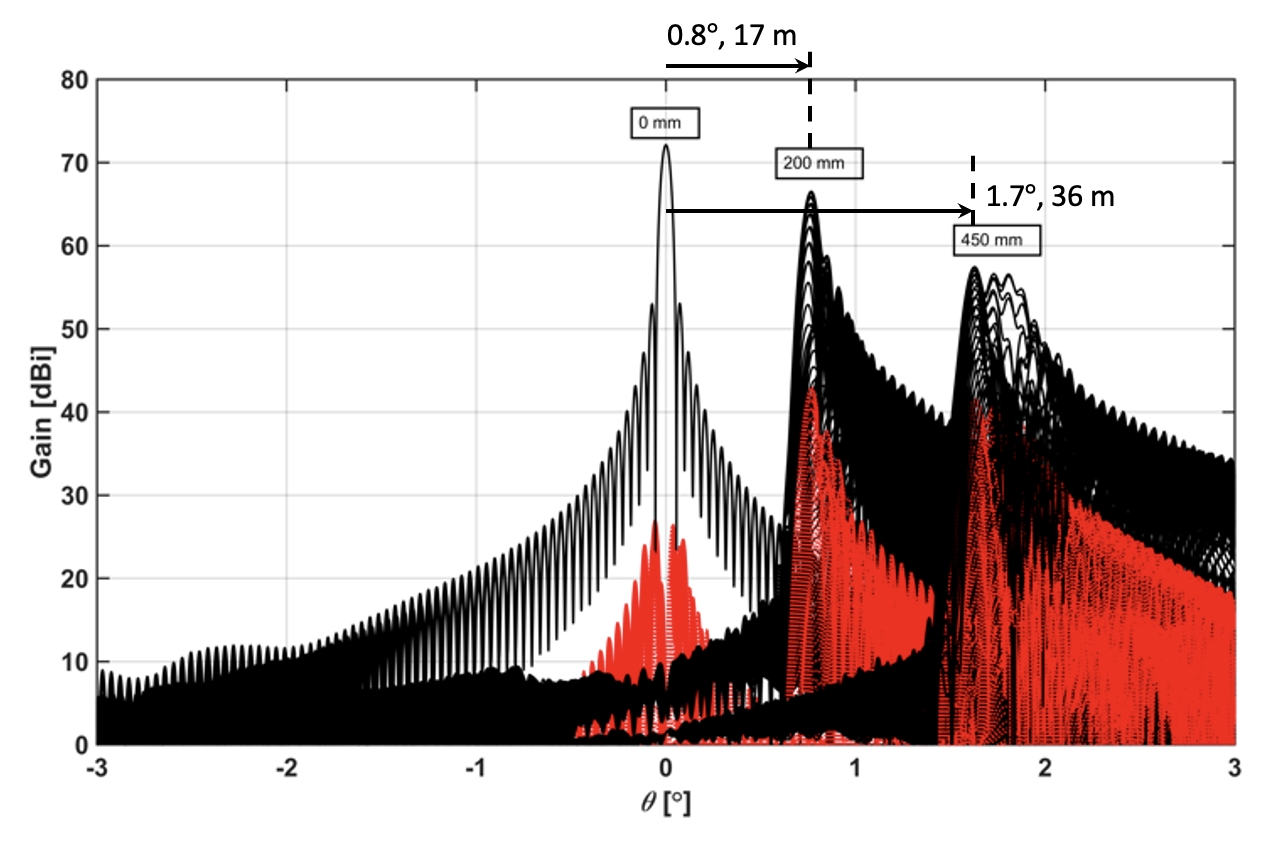}
\vspace{-0.3cm}
\caption{\small The Beam pattern of the Band 4 axially corrugated horn at 22 GHz on the ideal ngVLA antenna system. The pattern is shown over several azimuth angles for the horn offset by 0 mm, 200 mm, and 450 mm. The black and red lines are the co- and cross-polarized pattern cuts. The physical lengths at the tropospheric phase screen with an assumed indicative distance of 1.2 km are also marked (adapted from \cite{Lehmensiek2022WVROffsetBeamMemo}).}
\vspace{-5mm}
\label{Fig: OffsetBeams1D}
\end{center}
\end{figure} 
\section{Offset Beam Electromagnetic Analysis}

In order to gain quantitative understanding of the beam offsets, peak gains and beam efficiencies of the off-axis beams, we carried out EM analyses for the offset feed positions at 22 GHz \citep{Lehmensiek2022WVROffsetBeamMemo}. As shown in Fig 3, the resulting beam offsets are $\sim$0.8 and $\sim$1.7 deg, for 200 \& 450 mm feed offsets which translate to 17 m and 36 m linear distances at a height of 1.2 km. This is the approximate height scale at which the spatial structure function of the tropospheric phase fluctuations show a change to a shallower slope (slower growth with separation) at the VLA site, interpreted as marking the transition from 3-D to 2-D turbulence and generally taken as the representative distance to the phase screen \citep{1999RaSc...34..817C, 2020ngVLAMEM061}. This is consistent with independent estimate for the distance to the turbulent phase screen from the High Resolution Rapid Refresh (HRRR) forecast analysis from NOAA, where the atmospheric boundary layer is seen at a mean height  of $\sim$~1.2 km during winter months, the primary target period for Band 5 \& 6 observations (2024; Svoboda, 2025, private communication). 

Fig 4 shows the 2-D beam pattern for the 200 mm feed offset case. As expected, the offset beams are asymmetric, with beam efficiencies of 0.75 and 0.82 within 0.5 and 1 deg radius regions from the offset peak for the 200 mm case, the corresponding values for 450 mm being poor, at 0.44 and 0.51.

\begin{figure}[h!]
\begin{center}   
\includegraphics[width=0.7\linewidth]{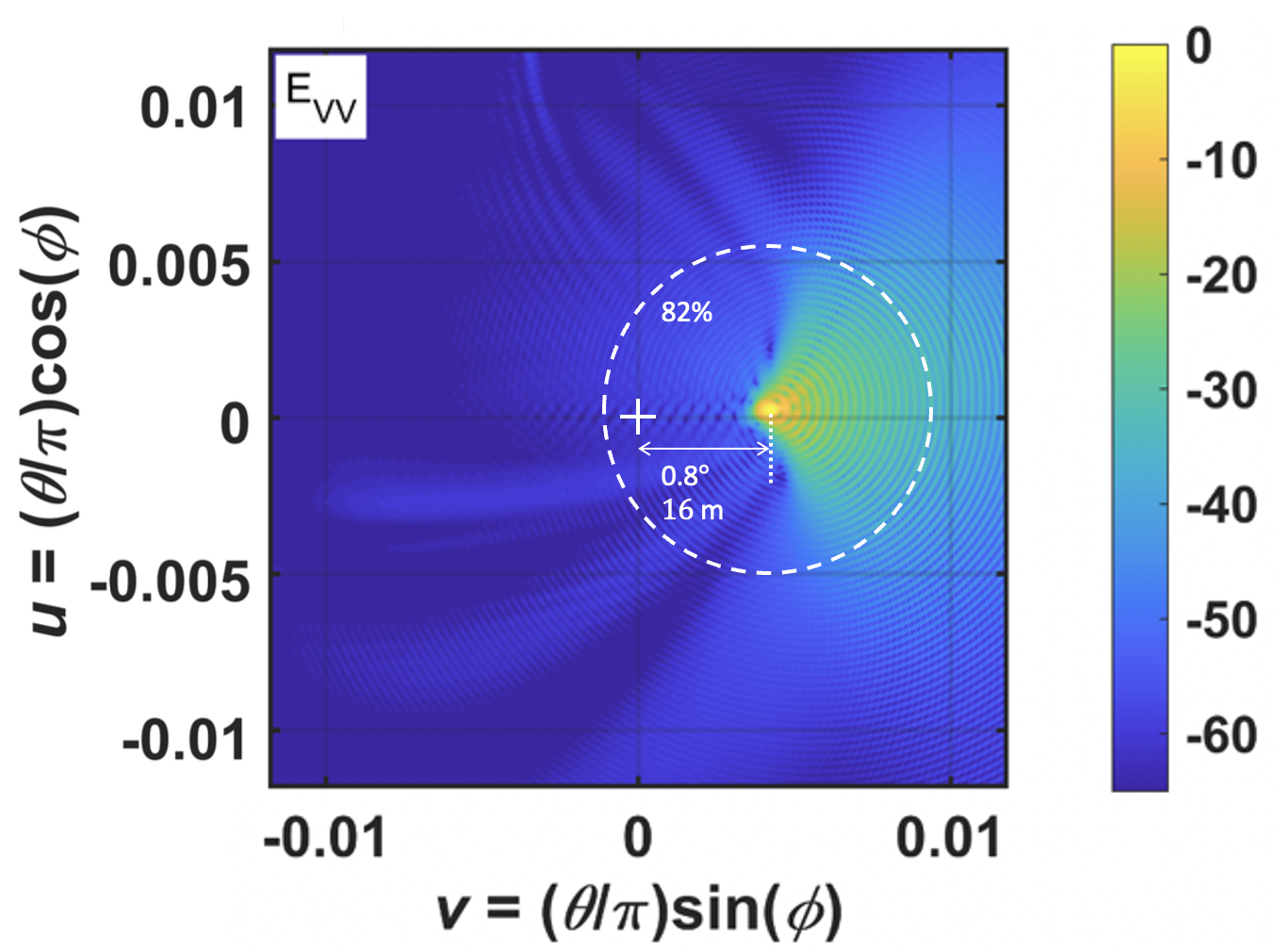}
%\vspace{-0.3cm}
\caption{\small: 2-D beam pattern at 22 GHz for 200 mm Band-4 feed offset. The 17 m distance for the 0.8$^{\circ}$ beam offset is for a tropospheric phase screen height of 1.2 km. The 82\% beam efficiency shown is for a circular region of 1$^{\circ}$ radius. The $+$ sign denotes the on-axis science beam direction. (adapted from \cite{Lehmensiek2022WVROffsetBeamMemo}).}
\vspace{-5mm}
\label{Fig: OffsetBeams2D}
\end{center}
\end{figure} 

\section{Offset Beam Implications, Other Limitations \& Mitigations}
We first note that the 0.8$^{\circ}$ angular offset for 200-mm feed offset is similar to the beam size of a standalone 1.5 m system and less than the angular distances to prospective fast switching reference calibrators. This means that the residual calibration errors realized with a Band-4 WVR should be similar or better.  The spatial offset at the phase screen for a 200 mm feed offset is $\sim$ one antenna diameter which is also the same as the length scale corresponding to a 4 second timescale for a tropospheric wind aloft of $\sim$ 10 m/s \citep{1999RaSc...34..817C}. Accordingly, we see the 200 mm offset case as a viable alternative capable of providing WVR data for Bands 5 \& 6 (and Band 4 as self-WVR). 

We will not discuss the 450 mm offset case further, given the above numbers, beyond pointing out that it can be used to estimate the bulk water vapor column at lower accuracy. Note that for this purpose, the asymmetric beam can still serve as a reliable probe of the bulk content as the tropospheric water emission will continue to be beam filling on a larger scale. 

In the general context of tropospheric delay calibration strategies, the Band-4 WVR concept can be thought of as fast switching to a calibrator 0.8° away, with zero cycling time (continuous). 
However, a minimum integration time is required to detect the water line which limits the minimum cycling time. The effective baseline that samples the tropospheric spatial phase structure function applicable in this situation is \citep{1999RaSc...34..817C}:
\begin{equation}
b_{eff} = v_a \times t_{cyc} / 2 + 0.014 \times h_{trop}
\end{equation}

where $v_a$ is the tropospheric wind aloft, $h_{trop}$ is the representative tropospheric water vapor height, $t_{cycl}$ is the WVR data acquisition sampling rate, and 0.014 is the 0.8$^{\circ}$ beam offset in radian.  Setting $h_{trop}$   to 1.2 km and $t_{cycl}$ to 4 s, $b_{eff}$ $\sim$ 40 m. In short, this concept renders the whole ngVLA baseline distribution to be contained within an effective baseline length of $\sim$ 40 m from a tropospheric delay fluctuation calibration perspective. 

We now turn to some shortcomings and present mitigating factors. While the comatic beam may appear undesirable at first sight, the beam efficiency within 1$^\circ$ is very good at 82\% (85\% with respect to the uncompromised beam efficiency of 96\% for an on-axis feed). More careful considerations suggest that the impact would be even less than implied by the beam efficiency. This is because a WVR scheme that tracks fluctuations, as opposed to total emission, is not as sensitive to beam efficiency. Considering  a 4 second time scale and a $v_a$ of 10 m/s, the corresponding tropospheric phase screen size scale from which the fluctuations arise is 20 m. 
Since we track the fluctuations of the water vapor line, we should consider the impact of the emission from the 18\% integrated beam response (= 100$-$82) of the error beam outside the 1 deg  region (= 20 m at 1.2 km) on the fluctuations. The contributions to fluctuations on $\sim$~4 s time scale from different $\sim$~20~m sized regions are independent and add incoherently, in quadrature. As an example, if we take the error beam to be composed of 10 $\times$ 1 deg regions,
then the incoherent combination of the fluctuations is 1/$\sqrt{10}$ of the contributions to the total emission. In this case, the beam efficiency for the fluctuating component would be \mbox{$\sim$ 94\% ( = 100 - 18/$\sqrt{10})$}. 

If found necessary, it will be possible to invoke a band-switching calibration strategy on a longer time scale in a layered approach to assess and possibly calibrate out the effect of the offset WVR beam. This would involve moving the Band-4 feed to the focus to obtain simultaneous calibrator and water vapor measurements with identical (selfsame) beams followed by water vapor measurements with the offset beam. This can be done on the 3-minute fast switching time scale or slower. If invoked, the 20-s ngVLA band switching time specification (SYS0908; \cite{2021ngVLASysReq}) will lower the 90\% calibration efficiency on 200-s time scale to 80\%. However, we expect such calibration on a celestial calibrator would only be needed on longer time scales  and the calibration efficiency can be maintained close to \mbox{$\sim$ 90\%}. The final cadence will have to be determined from field experience and can be tuned adaptively for individual sites and prevailing conditions \citep{2024ngVLACalConcept}. Note that such calibration is not possible with a standalone system which suffers both a spatial offset at the aperture and an angular offset as currently designed to intersect the main antenna boresight axis at a specific height.     

The increased ground pick up expected due to the offset primary illumination, which while increasing the $T_{sys}$, should only be a continuum component that can be mitigated by the large number of spectral channels. Additionally, it should also not fluctuate on seconds time scales and should therefore drop out in the differential WVR scheme. 
 %Further, in the near-field of the 18 m aperture whose Rayleig-Freznel distance at 22 GHz is well above the troposphere ($D^2/\lambda \sim$ 22 km), the response is likely to be better than 82\%. This can only be determined reliably with a near-field simulation as opposed to the far-field beam analyses we have obtained so far. In a simple approach we can model the near-field response of the science beam and the water vapor probe beam to be cylinder with their axial offset of 0.8 deg. Using the   
 
%The ngVLA Digital Back Ends (DBE), located in the antennas, provide spectral channelization at ~ 200 MHz resolution and cover the entire Band-4 passband of 20.5 -34 GHz. The WVR data acquisition would be through the DBEs and the necessary capability to construct the water vapor line spectra will need to be provided for the Band-4 DBEs. Such a capability is needed in any case for the stand alone WVR, in addition to dedicated DBEs. 

The parameters of the algorithm used to estimate the PWV can be calibrated using interferometric data from periodic calibrator visits both in the observing band of the science target (Band 5 or 6) and in Band 4 through band-switching. In current realizations, these parameters are  the channel weighting factors (e.g. \cite{2020ngVLAMEM074}). A more sophisticated approach incorporating radiative transfer in a model atmosphere applicable to specific antenna locations and time in a coarse grid, combined with short time scale, surface weather and on-antenna WVR data, is under development (Massingill et al 2026).

Finally, the frequency range covered around the water vapor line is limited by the Band-4 frequency span which extends down to 20.5 GHz, the upper end being 34 GHz. The standalone system does not suffer this limitation and can go to lower frequencies - 20 GHz in the current standalone baseline design \citep{2022WVRTechReq}. The impact of this limitation will be addressed by the ongoing path length retrieval algorithm study mentioned above (Massingill et al 2026). 

%\section{Advantages}

%The main advantage of this concept is the complete elimination of a separate WVR system, to be balanced against the drawbacks noted and their mitigating factors. The goal of this memo is to make an overall assessment on whether further studies are warranted to address finer details.

%An important technical advantage is that the WVR beam more closely samples the offending column fluctuating water vapor. This is because, the near-field response of the large primary aperture is a cylinder with an axial offset of 0.8° and the volume overlap with the on-axis cylinder applicable to the science observation is estimated to be ~ xx\% at 1.2 km compared to the ~ yy\% for the standalone WVR. With the bulk of the water vapor distribution being at lower heights along the column, the sampling is better than xx%.

\section {WVR System Requirements} 

The overall WVR excess path length measurement accuracy and timescale requirements based on detailed analysis and statistical modeling anchored by VLA site testing interferometer data have been presented in \cite{Asaki2025ngVLAMEM135}. The work to further break down these requirements into WVR hardware requirements (frequency coverage, number and widths of the spectral channels, sensitivity, and stability) and path length prediction algorithm accuracy requirements through tropospheric water line modeling is in progress (Massingil et al). Based on existing experience, we expect the prediction algorithm to be the weakest link. Pending ongoing WVR requirements break down work, we present below general tentative sensitivity and stability estimates assuming nominal parameters, to be updated when the above work is complete.
\\ \\
The requirements are:

\begin{itemize}
    \item The overall top level, implementation agnostic WVR system requirement is that it delivers improvements in post-correction residual fluctuations significantly better than 50\% of the time. This stems from the fact that at \mbox{$\sim$ 50\%} calibration overhead, the default fast switching strategy is available for phase corrections and substantially better WVR performance is required to justify the effort. A comparative cost analysis is needed to better define this requirement. The required WVR performance levels have not been demonstrated so far in an operational 22~GHz system, to our knowledge. However, in limited conditions (e.g. cloudless winter periods, preferred for observations in Bands 5 \& 6), we believe that the path length retrieval method being developed is a promising approach to enhance the success rate of achieving residual phase fluctuation improvements.
    
    \item The provision of a WVR digital data processor (same as or similar to the one for the standalone WVR).
     
\end{itemize}

The following requirements are specific to the Band-4 WVR: 

\begin{itemize}
    \item Configuring the feed layout to position the Band-4 feed at the center with Bands 5 \& 6 on either side.
    
    \item Ability to operate the Band-4 receiver simultaneously with another receiver (Band-5/6). This may include signal digitization, digital data transport, and processing similar to the standalone WVR with implementation design details driving the specific requirements (not addressed). 
    
\end{itemize}

\subsection{Sensitivity Estimates}

The raw thermal sensitivity required to obtain adequate measurement of the water vapor line is expected to be easily achieved with the Band-4 science receiver. We make estimates using the path length (or equivalently the PWV) measurement tolerance requirements derived in \cite{Asaki2025ngVLAMEM135}: excess path length accuracy of 0.17 - 0.14 mm for 1 - 4 second WVR correction timescales. This translates to 
peak differential line temperatures of 57 - 50 mK based on analytical estimates in \cite{Clarke2015WVRngVLAMEM10} (60 mK/0.18 mm of path length). Note that we are considering difference measurements here, where fluctuations of the water vapor line are tracked. This estimate is in excellent agreement with results in \cite{1999VLAMEMO177}, who obtains a peak difference line temperature  of $\sim$ 70 mK for the difference spectrum, with an excess precipitable water vapor content of 35 $\mu$m ($\sim$ 220 $\mu$m of excess path), through careful radiative transfer modeling.  We target 28 - 25 mK in order to detect the line at half this strength. Assuming a Band-4 system temperature of 32 K \citep{2024ngVLASysConDesRep} and a narrow 200 MHz channel width (compared to few GHz wide pressure broadened water vapor spectral line), the sensitivity for 1 - 4 seconds is 2.2 - 1.1 mK which readily provides 12 - 25 $\sigma$ detections. Allowing for a combined factor of 2 noise increase due to differential measurements, tracking only variations by differencing successive samples (spectra), and to account for the differencing between the two antennas of a baseline, 6 - 12 sigma measurements are feasible. In short, sufficient raw thermal sensitivity is available. Notably, for a standalone WVR with an ambient temperature receiver, the raw thermal sensitivity is not as easily met and would need longer time scales and/or wider channels. Based on the order magnitude larger $T_{sys}$ specified for the standalone WVR \citep{2022WVRTechReq} compared to Band-4 \citep{2024ngVLASysConDesRep} (480 K vs 32 K), it appears the $\Delta\nu \times \Delta t$ product would need to be $\times10^2$ higher for the standalone system to reach the same sensitivity. This difference can be narrowed, as much better ambient temperature noise figures can be realized but would still be near an order of magnitude.

\subsection{Stability Estimates}

In addition to thermal sensitivity, the WVR system stability is an important consideration in implementing corrections. Here again, as gain instabilities combine multiplicatively with the system noise, the lower system temperature of the Band-4 receiver allows a less stringent stability requirement compared to an ambient temperature standalone WVR system. This is a large relaxation, based on the $T_{sys}$ difference already noted for sensitivity. The stability timescale is set by the timescale for calibrating the WVR system. In this calibration, the parameters used to translate WVR line measurements to excess path shall be continuously updated to track changes due to the evolving states of the atmosphere and the WVR system, through periodic visits to a nearby calibrator. As previously noted, the ngVLA slew and settle time specifications allow 3-minute time scale for such calibration with 90\% calibration efficiency. Depending on the detected source flux levels, selfcal will become feasible at some timescale and can be applied to correct for longer term gain changes, including fluctuations of non-water vapor origin. 

With a Band-4 system temperature of 32 K, the estimated stability requirement is $< 8  \times$ 10$^{-4}$ on 3 minute time scales. The achievement of this stability is aided by a switched power diode noise cal system, which is an inherent part of the ngVLA Calibration concept and requirements \citep{2024ngVLACalConcept}. However, the WVR requirement is more stringent than the level needed for ngVLA relative amplitude calibration stability by a factor of $\sim$ 3 \citep{DraftCalReq}. Note that difference spectrum measurement would relax the requirement which can be leveraged if necessary. Better levels of stability using diode noise cal have already been demonstrated in a prototype Compact Water Vapor Radiometer (CWVR) \citep{2018arXiv180701690G}. It should also be pointed out that the timescale for this stability is much shorter than previous generation of WVRs developed at the VLA (e.g. \mbox{$\sim 10^3$~s} in CWVR). Additionally, there is little analog signal chain involved as the signal is digitized right in the receiver enclosure (e.g. the ngVLA Integrated Receivers and Digitizers (IRD) concept \citep{2024ngVLASysConDesRep, 2024ngVLAIRDSysConDesRep}), and no analog filters would be employed in constructing the water vapor spectra in the WVR digital processor. These factors place the Band-4 WVR in a much better position with respect to the exquisite thermal control otherwise usually needed in ambient temperature/standalone/analog systems.  

\section{CWVR Test Radiometers}

The functioning of the Compact Water Vapor Radiometers (CWVRs) currently installed (not operational) on 2 VLA antennas (2 others are in the lab undergoing tests) more closely resembles the Band-4 WVR concept than the small dish standalone WVR.  Their debugging and commissioning can provide a path for validating the Band-4 WVR concept. While the Band-4 WVRs can be expected to be superior to a functioning VLA K band based CWVR as they cover a larger bandwidth and provide a much finer spectral sampling, on-sky CWVR tests will be an important step forward in validating new prediction algorithms being developed and to obtain a statistical characterization of the water vapor delay fluctuation behavior at the VLA site. While results from other observatories with operational \mbox{22~GHz} WVR instruments can serve as pointers, direct data over extended periods for the VLA site are critical for careful assessments.  

\section{Conclusion}

We have developed and studied a WVR concept  for the ngVLA which utilizes the standard Band-4 science receivers and the main antennas as an alternative to the current baseline concept consisting of a large number of dedicated, standalone $\sim$1.5 m antenna based systems with their own receivers and electronics, one unit at each antenna. This included an EM analysis of the characteristics of the offset beam, its overlap with the science beam,  and considerations of spatial and time scales of water vapor fluctuations sensed, leading to the conclusion that adequate performance would be feasible with the Band-4 system. Besides eliminating a separate  WVR subsystem, the Band-4 WVR concept also offers some advantages over the standalone system. 

The Band-4 WVR concept requires placement of the Band-4 feed at the center with Bands 5 \& 6 on either side and the ability to simultaneously operate two receivers and associated signal chains. 

The work on the CWVR test radiometers should be prioritized to derisk the WVR correction strategy by enabling more direct and complete characterization of path length fluctuations at the VLA site and through the verification of new path length retrieval algorithms. 

Based on this study we recommend that the implementation details of adopting the Band-4 WVR concept as the baseline design for the ngVLA digital water vapor radiometer (DWVR) be carefully considered. Adopting the Band 4 WVR concept does not preclude a standalone system in future if needed, whereas front end and digital data transport design choices can exclude the Band 4 concept.

\section*{Acknowledgement}

We thank Chris Carilli, Kyle Massingil, Rob Selina, Brian Svoboda, and Jeff Mangum for reading and commenting on the manuscript which led to improvements. This memo benefited from discussions with Sanjay Bhatnagar and Wes Grammer. We gratefully acknowledge Brian Svoboda's help in providing the HRRR/NOAA data for the height of the planetary boundary layer. 
\\ \\
{\it The National Radio Astronomy Observatory and Green Bank Observatory are facilities of the U.S. National Science Foundation operated under cooperative agreement by Associated Universities, Inc. This work was supported by awards AST-2034328 (MSIP Prototype Antenna) and AST-2334267 (ngVLA Design Activities); NRAO related activities are funded under award AST-1647378 (NRAO Operations/Development).}

\begin{figure}[h!]
\begin{center}   

\includegraphics[width=\textwidth]{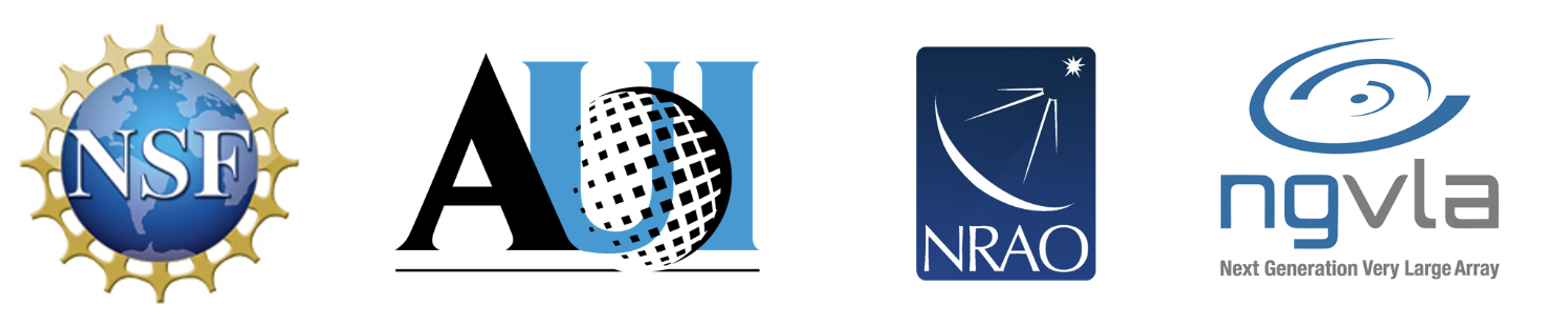}

\end{center}
\end{figure} 

% References
\bibliography{PASPsample631}{}
%\bibliography{ngVLA-230G}{}
\bibliographystyle{aasjournal}

\end{document}